\documentclass[letterpaper,preprintnumbers,prd,twocolumn,nofootinbib,nobibnotes,showpacs]{revtex4}
\usepackage{epsfig}
\usepackage{graphicx}%
\usepackage{dcolumn}
\usepackage{amsmath}

\makeatletter
\def\btt#1{\texttt{\@backslashchar#1}}%
\DeclareRobustCommand\bblash{\btt{\@backslashchar}}%
\makeatother

\begin{document}

\title{Black Holes in Brans-Dicke Theory with a Cosmological Constant}% Force line breaks with \\
\author{Chang Jun Gao$^1$}\email{gaocj@mail.tsinghua.edu.cn}\author{Shuang Nan Zhang$^{1,2,3,4}$}
\email{zhangsn@mail.tsinghua.edu.cn}\affiliation{$^1$Department of
Physics and Center for Astrophysics, Tsinghua University, Beijing
100084, China(mailaddress)} \affiliation{$^2$Physics Department,
University of Alabama in Huntsville, AL 35899, USA}
\affiliation{$^3$Space Science Laboratory, NASA Marshall Space
Flight Center, SD50, Huntsville, AL 35812, USA}
\affiliation{$^4$Key Laboratory of Particle Astrophysics,
Institute of High Energy Physics, Chinese Academy of Sciences,
Beijing 100039, China}

\date{\today}% It is always \today, today, but you may specify any date with \date.

\begin{abstract}
Since the Brans-Dicke theory is conformal related to the dilaton
gravity theory, by applying a conformal transformation to the
dilaton gravity theory, we derived the cosmological constant term
in the Brans-Dicke theory and the physical solution of black holes
with the cosmological constant. It is found that, in four
dimensions, the solution is just the Kerr-Newman-de Sitter
solution with a constant scalar field. However, in $n>4$
dimensions, the solution is not yet the $n$ dimensional
Kerr-Newman-de Sitter solution and the scalar field is not a
constant in general. In Brans-Dicke-Ni theory, the resulting
solution is also not yet the Kerr-Newman-de Sitter one even in
four dimensions. The higher dimensional origin of the Brans-Dicke
scalar field is briefly discussed.
\end{abstract}

 \maketitle
\section{Introduction}
In 1915, Einstein constructed the relativistic theory of gravity,
i.e. general relativity. General relativity is extremely
successful at describing the dynamics of our solar system and
perhaps the observable Universe. However, general relativity
probably does not describe gravity accurately at all scales. Two
kinds of problems coming from two different ways, namely that, the
singularity problem at small scales and the dark energy problem at
large scales, may account for this point explicitly. Since general
relativity is also a classical theory, it is natural that it will
face the singularity problem. As regards dark energy, Lovelock
showed us long ago that it was an entirely natural part of general
relativity, namely, the cosmological constant term being the dark
energy. However, if we take the cosmological constant term as the
dark energy, we will face two even harder problems, the well-known
cosmological problem and the correspondence problem. It is widely
believed that these problems may be overcome in the quantum
gravity. On the other hand, general relativity does not
accommodate either Mach's principle or Dirac's large-number
hypothesis. Herein, various alternatives of gravity have been
explored. \\
\hspace*{3.5mm}As the simplest modification of general relativity,
Brans and Dicke developed another relativistic theory of gravity,
namely that, the notable Brans-Dicke theory \cite{1}. Compared to
Einstein's general relativity, Brans-Dicke theory describes the
gravitation in terms of the metric as well as a scalar field and
accommodates both Mach's principle and Dirac's large-number
hypothesis as its new ingredients. Furthermore, Brans-Dicke theory
also passed all the available observational and experimental tests
\cite{2}. Unfortunately, the singularity problem remains in this
theory.\\
\hspace*{3.5mm}Since the gravitational collapse and the subsequent
black hole formation is of great importance in classical gravity,
many authors have investigated these aspects in Brans-Dicke theory
\cite{3}. Hawking has proved that in four dimensions, the
stationary and vacuum Brans-Dicke solution is just the Kerr
solution with constant scalar field everywhere \cite{4}. Cai and
Myung have proved that in four dimensions, the charged black hole
solution in the Brans-Dicke-Maxwell theory is just the
Reissner-Nordstr$\ddot{\textrm{o}}$m solution with a constant
scalar field \cite{5}. Since one had no knowledge of the
cosmological constant term in the Brans-Dicke theory, one had not
obtained the solution of black holes with the cosmological
constant. Thus the goal of this paper is to report we have found
the explicit expression of the cosmological term and the solution
of de Sitter version for black holes in Brans-Dicke theory.
\section{black holes in four dimensions}
We start from the actions of dilaton gravity theory and the
Brans-Dicke theory in four dimensions which are given by
\begin{eqnarray}
\bar{S}&=&\int{d^4x\sqrt{-\bar{g}}}\left[\bar{R}-{2}\bar{\nabla}_{\mu}\bar{\phi}\bar{\nabla}^{\mu}{\bar{\phi}}
-\bar{V}\left(\bar{\phi}\right)\right.\nonumber
\\ &&\left.  -e^{-{2\alpha\bar{\phi}}}\bar{F}^2\right],\nonumber\\
{S}&=&\int{d^4x\sqrt{-{g}}}\left[{\phi
R}-{\frac{\omega}{\phi}}{\nabla}_{\mu}{\phi}{\nabla}^{\mu}{{\phi}}
-V\left({\phi}\right)\right.\nonumber
\\ &&\left.  -{F}^2\right],
\end{eqnarray}
where $R, \bar{R}$ are the scalar curvature, $F^2,\bar{F}^2$ are
the usual Maxwell contribution, $\alpha$ is an arbitrary constant
governing the strength of the coupling between the dilaton and the
Maxwell field, and $V\left(\bar{\phi}\right)$ is the potential of
dilaton which is with respect to the cosmological constant given
by \cite{6}
\begin{eqnarray}
\bar{V}\left(\bar{\phi}\right)&=&\frac{2\lambda}{3\left(1+\alpha^2\right)^2}\left[\alpha^2\left(3\alpha^2-
1\right)e^{-2\bar{\phi}/\alpha}
\right.\nonumber\\&&\left.+\left(3-\alpha^2\right)e^{2\bar{\phi}\alpha}+8\alpha^2e^{\bar{\phi}\alpha-\bar{\phi}/\alpha}\right].
\end{eqnarray}
Here $\lambda$ is the Einstein's cosmological constant. When
$\alpha=0$ or $\phi=0$, the potential reduces to the usual
Einstein cosmological constant. As for the scalar potential
$V(\phi)$, we have no knowledge about it. In the next, we would
derive this potential by performing a conformal transformation.\\
\hspace*{3.5mm}Varying the actions with respect to the scalar
fields, respectively, yields
\begin{eqnarray}
\bar{\nabla}^2\bar{\phi}=\frac{1}{4}\frac{d \bar{V}}{d
\bar{\phi}}-\frac{\alpha}{2}e^{-2\bar{\phi}}\bar{F}^2,\nonumber\\
{\nabla}^2{\phi}=\frac{1}{2\omega+3}\left(\phi\frac{d
V}{d{\phi}}-2V\right).
\end{eqnarray}
Using equations (3), if we perform a conformal transformation with
\begin{eqnarray}
\bar{g}_{\mu\nu}=\phi {g}_{\mu\nu},
\end{eqnarray}
and demand
\begin{eqnarray}
\sqrt{-\bar{g}}\bar{\nabla}_{\mu}\bar{\phi}\bar{\nabla}^{\mu}{\bar{\phi}}&\propto&
\sqrt{-{g}}\frac{1}{\phi}{\nabla}_{\mu}{\phi}{\nabla}^{\mu}{{\phi}},\nonumber\\
\sqrt{-\bar{g}}e^{-{2\alpha\bar{\phi}}}\bar{F}^2&\propto&\sqrt{-{g}}{F}^2,
\end{eqnarray}
we get the following relations
\begin{eqnarray}
\alpha=0, \ \ \ \ \bar{\phi}=f_0\left(\ln\phi-\ln\phi_0\right),
\end{eqnarray}
where $f_0$ and $\phi_0$ are two integration constants. The
resulting action of Brans-Dicke theory is
\begin{eqnarray}
\bar{S}=S&=&\int{d^4x\sqrt{-{g}}}\left[{\phi
R}-{\frac{\omega}{\phi}}{\nabla}_{\mu}{\phi}{\nabla}^{\mu}{{\phi}}
-{F}^2\right.\nonumber\\&&\left.-\left(2\lambda\phi^2+\Lambda\phi^{3+\frac{2}{3}\omega}\right)\right],
\end{eqnarray}
where $\Lambda$ is an integration constant and
$\omega=2f_0^2-{3}/{2}$. The cosmology with the potential
$V(\phi)=2\lambda\phi^2$ have been discussed by Santos and Gregory
\cite{7}. Torres have investigated the cosmology with even more
general potential $V=\phi^{n}$ \cite{8}. Here we
have derived it theoretically.\\
\hspace*{3.5mm}We see that when $\alpha=0$, the dilaton gravity
theory reduces to the Einstein-Maxwell-scalar theory and the
physical black hole solution is just the Kerr-Newman-de Sitter
solution with a constant scalar field $\bar{\phi}$. On the other
hand, Eq.(6) indicates that $\phi$ is also a constant. Thus we see
that the physical black hole solution in Brans-Dicke theory in
four dimensions is still the Kerr-Newman-de Sitter one. In this
aspect, Hawking has proved that in the four dimensional vacuum
Brans-Dicke theory, the black hole solution is just the Kerr
solution with a constant scalar field. Cai and Myung have proved
that in four dimensions, the charged black hole solution in the
Brans-Dicke-Maxwell theory is just the
Reissner-Nordstr$\ddot{\textrm{o}}$m solution with a constant
scalar field. Thus our conclusion is an extension of their
results. We note that the constant $\Lambda$ is not present in the
solution. Thus the term $2\lambda\phi^2$ in Eq.(7) corresponds to
the Einstein cosmological constant. It is a remarkably simple
expression.\\
\section{black holes in higher dimensions}
The higher dimensional analogue of the actions for the dilaton
gravity theory and the Brans-Dicke theory are
\begin{eqnarray}
\bar{S}&=&\int{d^nx\sqrt{-\bar{g}}}\left[\bar{R}-\frac{4}{n-2}\bar{\nabla}_{\mu}\bar{\phi}\bar{\nabla}^{\mu}{\bar{\phi}}
-\bar{V}\left(\bar{\phi}\right)\right.\nonumber
\\ &&\left.  -e^{-{\frac{4\alpha\bar{\phi}}{n-2}}}\bar{F}^2\right],\nonumber\\
{S}&=&\int{d^nx\sqrt{-{g}}}\left[{\phi
R}-{\frac{\omega}{\phi}}{\nabla}_{\mu}{\phi}{\nabla}^{\mu}{{\phi}}
-V\left({\phi}\right)\right.\nonumber
\\ &&\left.  -{F}^2\right],
\end{eqnarray}
where \cite{9}
\begin{eqnarray}
&&\bar{V}\left(\bar{\phi}\right)=\frac{\lambda}{3\left(n-3+\alpha^2\right)^2}
\nonumber\\&&\cdot\left[-\alpha^2\left(n-2\right)\left(n^2-n\alpha^2-6n+\alpha^2+9\right)
e^{-\frac{4\left(n-3\right)\bar{\phi}}{\left(n-2\right)\alpha}}\right.\nonumber
\\ &&\left. +\left(n-2\right)
\left(n-3\right)^2\left(n-1-\alpha^2\right)e^{\frac{4\alpha\bar{\phi}}{n-2}}\right.\nonumber
\\ &&\left. +4\alpha^2\left(n-3\right)
\left(n-2\right)^2e^{\frac{-2\bar{\phi}\left(n-3-\alpha^2\right)}{\left(n-2\right)\alpha}}\right].
\end{eqnarray}
\hspace*{3.5mm}Varying the actions with respect to the scalar
fields, respectively, yields
\begin{eqnarray}
&&\bar{\nabla}^2\bar{\phi}=\frac{n-2}{8}\frac{d \bar{V}}{d
\bar{\phi}}-\frac{\alpha}{2}e^{-\frac{4\alpha\bar{\phi}}{n-2}}\bar{F}^2,\nonumber\\
&&{\nabla}^2{\phi}=-\frac{n-4}{2\left[n-1+\omega\left(n-2\right)\right]}F^2
\nonumber\\&&+\frac{1}{2\left[n-1+\omega\left(n-2\right)\right]}\left[\left(n-2\right)\phi\frac{d
V}{d{\phi}}-nV\right].
\end{eqnarray}
Using equations (10), if we perform a conformal transformation
with
\begin{eqnarray}
\bar{g}_{\mu\nu}=\phi^{\frac{n}{n-2}} {g}_{\mu\nu},
\end{eqnarray}
and demand
\begin{eqnarray}
\sqrt{-\bar{g}}\bar{\nabla}_{\mu}\bar{\phi}\bar{\nabla}^{\mu}{\bar{\phi}}&\propto&
\sqrt{-{g}}\frac{1}{\phi}{\nabla}_{\mu}{\phi}{\nabla}^{\mu}{{\phi}},\nonumber\\
\sqrt{-\bar{g}}e^{-\frac{4\alpha\bar{\phi}}{n-2}}\bar{F}^2&\propto&\sqrt{-{g}}{F}^2,
\end{eqnarray}
we get the following relations
\begin{eqnarray}
\bar{\phi}=\frac{n-4}{4\alpha}\left(\ln\phi-\ln\phi_0\right),
\end{eqnarray}
where $\phi_0$ is an integration constant. The resulting action of
Brans-Dicke theory is given by
\begin{eqnarray}
\bar{S}=S&=&\int{d^nx\sqrt{-{g}}}\left[{\phi
R}-{\frac{\omega}{\phi}}{\nabla}_{\mu}{\phi}{\nabla}^{\mu}{{\phi}}
-{F}^2\right.\nonumber
\\ &&\left. -V\left(\phi\right)\right],
\end{eqnarray}
where
\begin{eqnarray}
&&\omega=\frac{\left(n-4\right)^2}{4\alpha^2\left(n-2\right)}-\frac{n-1}{n-2},\nonumber\\
&&V=\frac{1}{3}\lambda\phi_0^{\frac{n}{n-2}}\frac{\left(n-2\right)^2\left(n-4\right)}{\left(n-3+\alpha^2\right)^2}
\nonumber\\
&&\cdot\left[\frac{\left(n-3\right)^2\left(n-1-\alpha^2\right)}{n^2-4n\alpha^2-6n+8+4\alpha^2}
\exp{\frac{8\alpha\bar{\phi}}{n-4}}\right.\nonumber
\\ &&\left.+\frac{\alpha^2\left(-n^2+6n-\alpha^2+n\alpha^2-9\right)}{5n^2-22n+20}\right.\nonumber
\\ &&\left.\cdot\exp{\frac{4\bar{\phi}\left(n\alpha^2-n^2+7n-12\right)}{\left(n-4\right)
\left(n-2\right)\alpha}}\right.\nonumber
\\ &&\left.+\frac{4\alpha^2\left(n^2-5n+6\right)}{3n^2-2n\alpha^2-14n+2\alpha^2+14)}\right.\nonumber
\\ &&\left.\cdot\exp{\frac{2\bar{\phi}\left(3n\alpha^2-n^2+7n-4\alpha^2-12\right)}
{\left(n^2-6n+8\right)\alpha}}\right] \nonumber\\&&+\Lambda
\exp{\frac{\left(-4n\alpha^2+4\alpha^2n^2+32n-32-10n^2+n^3\right)\bar{\phi}}{\left(n-1\right)
\left(n-2\right)\left(n-4\right)\alpha}}, \nonumber\\&&
\phi_0=\left\{1+\frac{n^2-5n+4}{\left(n-2\right)\left[n-1+\omega\left(n-2\right)\right]}\right\}^{\frac{n-2}{n-4}}.
\end{eqnarray}
$\Lambda$ is an integration constant. For simplicity in writing,
the potential is expressed as the function of $\bar{\phi}$. At
first glance, it is very complicated. In fact, the potential
consists of simply four monomials of the scalar field $\phi$.\\
\hspace*{7.5mm}The higher dimensional non-rotating dilaton black
hole solution $(\bar{g}_{\mu\nu},\bar{\phi},\bar{F}_{\mu\nu})$
with the cosmological constant has been constructed by us
\cite{9}. By applying the conformal transformation in Eq.(11) and
Eq.(13), we are easy to write out the higher dimensional one with
the cosmological constant in Brans-Dicke theory. In general, the
resulting metric is not identical to the
Reissner-Nordstr$\ddot{\textrm{o}}$m-de Sitter solution. In four
dimensions, since the scalar field is constant everywhere, if a
massive body collapses behind an event horizon, its effect as a
source of the scalar field decreases to zero. So Hawking concluded
that there will not be any scalar gravitational radiation emitted
when two black holes collide in the four dimensional Brans-Dicke
theory. In higher dimensions, the scalar field is not constant
everywhere, so scalar gravitational radiation will occur when
black holes collide. We note again that $\Lambda$ does not present
in the black hole solution. Thus the $\Lambda$ term in the
potential exerts no influences on the properties
 of the black hole. However, recalling the discussions of Santos and
Gregory, and Torres, we conclude that the $\Lambda$ term makes
great contributions to the evolution of the Universe.
Here the $\lambda$ term in the potential is the counterpart of the Einstein cosmological constant.\\
\section{black holes in Brans-Dicke-Ni theory}
\hspace*{3.5mm}In section \textrm{I}, we find that if and only if
the coupling $\alpha=0$, the dilaton gravity theory reduces to the
Brans-Dicke theory via conformal transformation. In this section,
we show that when $\alpha\neq 0$, the dilaton gravity theory
reduces to the Brans-Dicke-Ni theory via conformal transformation.
The action of the four dimensional Brans-Dicke-Ni theory is given
by
\begin{eqnarray}
{S}&=&\int{d^4x\sqrt{-{g}}}\left[{\phi
R}-{\frac{\omega}{\phi}}{\nabla}_{\mu}{\phi}{\nabla}^{\mu}{{\phi}}
-V\left({\phi}\right)\right.\nonumber
\\ &&\left.  +f\left(\phi\right){F}^2\right],
\end{eqnarray}
where
\begin{eqnarray}
{\pounds}&=&\sqrt{-g}f\left(\phi\right){F}^2,
\end{eqnarray}
is the Lagrangian density found by Ni \cite{10}. In general, $f$
is a scalar function of other fields which includes scalar, vector
and any other fields in the theory of gravity under consideration.
However, in Brans-Dicke-Ni theory, $f$ is only the function of
Brans-Dicke scalar field. Carroll and Field have investigated
following Brans-Dicke-Ni action \cite{11}
\begin{eqnarray}
{S}&=&\int{d^4x\sqrt{-{g}}}\nonumber
\\ &&\cdot\left[{\phi
R}-{\frac{\omega}{\phi}}{\nabla}_{\mu}{\phi}{\nabla}^{\mu}{{\phi}}
-\left(1+\beta\phi\right){F}^2\right].
\end{eqnarray}
$\beta$ is some coupling constant. The Brans-Dicke-Ni theory leads
to the violation of the
Einstein equivalence principle while obeying the weak equivalence principle.  \\
\hspace*{3.5mm}Varying the action in Eq.(16) with respect to the
scalar field yields
\begin{eqnarray}
{\nabla}^2{\phi}=\frac{-1}{2\omega+3}\phi\frac{df}{d\phi}F^2+\frac{1}{2\omega+3}\left(\phi\frac{\partial
V}{\partial{\phi}}-2V\right).
\end{eqnarray}
Using equation (19), if we perform a conformal transformation in
the action $\bar{S}$ in Eqs.(1) with
\begin{eqnarray}
\bar{g}_{\mu\nu}=\phi {g}_{\mu\nu}, \ \ \ \
\bar{\phi}=f_0\left(\ln\phi-\ln\phi_0\right),
\end{eqnarray}
where $f_0, \phi_0$ are constants, we find that the dilaton
gravity theory becomes the Brans-Dicke-Ni theory
\begin{eqnarray}
\bar{S}=S&=&\int{d^4x\sqrt{-{g}}}\left[{\phi
R}-{\frac{\omega}{\phi}}{\nabla}_{\mu}{\phi}{\nabla}^{\mu}{{\phi}}
-V\left({\phi}\right)\right.\nonumber
\\ &&\left.  +f\left(\phi\right){F}^2\right],
\end{eqnarray}
where
\begin{eqnarray}
&&\omega=2f_0^2-\frac{3}{2},\nonumber\\
&&f\left(\phi\right)=f_1\phi^{1+2\omega/3}-\frac{1+2\omega/3}{1+2\omega/3-2\alpha f_0}\phi^{2\alpha f_0},\nonumber\\
&&V=\Lambda \phi^{3+2\omega/3}+\frac{4\lambda
f_0}{3\left(1+\alpha^2\right)^2}
\nonumber\\
&&\cdot\left[\frac{\alpha^3\left(3\alpha^2-1\right)\phi^{2-2f_0/\alpha}\phi_0^{2f_0/\alpha}}{\left(2f_0\alpha+3\right)}\right.\nonumber
\\ &&\left.+\frac{\left(\alpha^2-3\right)\phi^{2+2f_0\alpha}\phi_0^{-2f_0\alpha}}{\left(3\alpha-2f_0\right)}\right.\nonumber
\\ &&\left.-\frac{16\alpha^3\phi^{2+f_0\alpha-f_0/\alpha}\phi_0^{f_0/\alpha-f_0\alpha}}{\left(3\alpha^2-4f_0\alpha-3\right)}\right].
\end{eqnarray}
$f_1, \Lambda$ are two integration constants. The potential also
consists of four monomials of the scalar field $\phi$ and when
$\alpha=0, \phi_0=1$, it reduces to the potential in Eq.(7). Four
dimensional rotating  \cite{12}  and Non-rotating  \cite{13}
dilaton black hole solutions without the cosmological constant
have been found by Sen and Gibbons etc. Via conformal
transformation, we are easy to write out their version in the
presence of the cosmological constant in the Brans-Dicke-Ni
theory. From the resulting solutions, we can conclude that the
black hole solution in Brans-Dicke-Ni theory is not yet
Kerr-Newman-de Sitter one and $\lambda$ term in the potential is
with respect to the Einstein cosmological constant. $\Lambda$ term
exerts no influences on the properties
 of the black hole but would make contributions to the evolution of the Universe.
\section{higher dimensional origin of Brans-Dicke scalar field}
\hspace*{3.5mm}In the previous sections we have verified that the
Brans-Dicke theory is conformal related to the dilaton gravity
theory and the corresponding cosmological constant term is
derived. In this section, we discuss the higher dimensional origin
of the four dimensions Brans-Dicke scalar field. Our discussion is
motivated by the elegant
work of Gibbons et al \cite{14}.\\
\hspace*{3.5mm}Consider the $(4+p+q)$ dimensional vacuum Einstein
theory
\begin{eqnarray}
{S_{4+p+q}}&=&\int{d^{4+p+q}x\sqrt{-{g_{4+p+q}}}}R_{4+p+q}.
\end{eqnarray}
Assume the $(4+p+q)$ dimensional metric takes the form of
\begin{eqnarray}
{ds_{4+p+q}^2}&=&e^{2\alpha\psi\left(x\right)}dy\cdot dy
+e^{2\gamma\psi\left(x\right)}dz\cdot
dz\nonumber\\&&+e^{2\beta\psi\left(x\right)}g_{\mu\nu}dx^{\mu}dx^{\nu}.
\end{eqnarray}
We note that the metric is only depend on the coordinates of the
four dimensional submanifold. Then the $(4+p+q)$ dimensional
action reduces to the four dimensional one
\begin{eqnarray}
{S}&=&\int{d^{4}x\sqrt{-{g}}}e^{\left(p\alpha+q\gamma+2\beta\right)\psi}\left[R\right.\nonumber
\\ &&\left.-2\left(3\beta+p\alpha+q\gamma\right)\nabla^{2}\psi\right.\nonumber
\\ &&\left.-\left(6\beta^2+p^2\alpha^2+q^2\gamma^2+p\alpha^2+q\gamma^2
\right.\right.\nonumber
\\ &&\left.\left.+4q\gamma\beta+4p\alpha\beta+2p\alpha q\gamma\right)\nabla_{\mu}\psi\nabla^{\mu}\psi\right].
\end{eqnarray}
Put
\begin{eqnarray}
e^{\left(p\alpha+q\gamma+2\beta\right)\psi}&=&\phi,
\end{eqnarray}
then the action reduces to
\begin{eqnarray}
&&{S}=\int{d^{4}x\sqrt{-{g}}}\left[\phi
R-\frac{6\beta+2p\alpha+2q\gamma}{2\beta+p\alpha+q\gamma}\nabla^{2}\phi\right.\nonumber
\\ &&\left.-\left(p\alpha^2+
q\gamma^2-6\beta^2-p^2\alpha^2-q^2\gamma^2-6q\gamma\beta-6\beta
p\alpha\right.\right.\nonumber
\\ &&\left.\left.-2p\alpha q\gamma\right)\cdot
{(p\alpha+q\gamma+2\beta)^{-2}\phi^{-1}}\nabla_{\mu}\phi\nabla^{\mu}\phi\right].
\end{eqnarray}
On the other hand, we are aware that the Brans-Dicke theory is
given by
\begin{eqnarray}
{S}&=&\int{d^4x\sqrt{-{g}}}\left[{\phi
R}-{\frac{\omega}{\phi}}{\nabla}_{\mu}{\phi}{\nabla}^{\mu}{{\phi}}
-V\left({\phi}\right)\right],
\end{eqnarray}
from which we have the equation of motion
\begin{eqnarray}
{\nabla}^2{\phi}=\frac{1}{2\omega+3}\left(\phi\frac{d
V}{d{\phi}}-2V\right).
\end{eqnarray}
Substituting Eq.(29) into Eq.(27) and setting
\begin{eqnarray}
\omega &=&\left(p\alpha^2+
q\gamma^2-6\beta^2-p^2\alpha^2-q^2\gamma^2-6q\gamma\beta\right.\nonumber
\\ &&\left.-6\beta
p\alpha-2p\alpha q\gamma\right)\cdot
{(p\alpha+q\gamma+2\beta)^{-2}},\nonumber\\
V&=&\frac{6\beta+2p\alpha+2q\gamma}{2\beta+p\alpha+q\gamma}\cdot\frac{1}{2\omega+3}\left(\phi\frac{d
V}{d{\phi}}-2V\right),
\end{eqnarray}
we can find that when
\begin{eqnarray}
p\alpha+q\gamma=0,
\end{eqnarray}
Eq.(28) turns out to be
\begin{eqnarray}
{S}&=&\int{d^4x\sqrt{-{g}}}\left[{\phi
R}-{\frac{\omega}{\phi}}{\nabla}_{\mu}{\phi}{\nabla}^{\mu}{{\phi}}
-\Lambda\phi^{3+\frac{2}{3}w}\right].
\end{eqnarray}
$\Lambda$ is an integration constant. It is exactly the action of
Eq.(7) in the absence of the cosmological constant. Thus we see
that both Brans-Dicke scalar field and $\Lambda$ term originate
from the extra dimensions of the vacuum Einstein theory.
\section{conclusion and discussion}
\hspace*{3.5mm}In conclusion, by applying the conformal
transformations, we have derived the cosmological constant term in
the Brans-Dicke theory. It is found that, in four dimensions, the
resulting black hole solution is just the Kerr-Newman-de Sitter
solution in general relativity. However, this is not the truth for
higher dimensions. On the other hand, in Brans-Dicke-Ni theory,
even in four dimensions, the resulting black hole solution is not
yet the Kerr-Newman-de Sitter one. We had better point out that in
Brans-Dicke theory, the scalar potential consists of two terms one
of which, $\lambda$ term, is for the Einstein cosmological
constant and the other, $\Lambda$ term, has no counterpart in
general relativity. $\lambda$ term exerts influences on both the
properties of the black hole and the evolution of large scale
Universe. In contrast, $\Lambda$ term exerts no influences on the
properties
 of the black hole. But the investigations of some specific forms of this term
\cite{7,8} have revealed that it plays some important role in the
evolution of the Universe. Thus we should not omit it freely. It
is found that both the Brans-Dicke scalar field and the $\Lambda$
term can be viewed as the contributions of extra dimensions of the
vacuum Einstein theory.
\begin{acknowledgments}
\hspace*{3.5mm}This study is supported in part by the Special
Funds for Major State Basic Research Projects, by the Directional
Research Project of the Chinese Academy of Sciences and by the
National Natural Science Foundation of China. SNZ also
acknowledges supports by NASA's Marshall Space Flight Center and
through NASA's Long Term Space Astrophysics Program.
\end{acknowledgments}

\end{document}